\documentclass{article}

\usepackage{graphicx}

\usepackage{color}
\usepackage{epstopdf}

\begin{document}
\bibliographystyle{unsrt}

\title{Transient gain from  $N_2^+$ in light filaments}%

\author{Ladan Arissian$^{1,2}$, Brian Kamer$^1$, Ali Rastegari$^1$,\\ 
\vspace{5mm}
  David Villeneuve$^{2,3}$and Jean Claude Diels$^1$\\
\small 1. Center for High Technology Materials, Albuquerque, NM, USA\\
\small 2. Department of Physics, University of Ottawa, Ottawa, Canada \\
\small 3. Joint Attosecond Science Laboratory, National Research Council of Canada, Ottawa, Canada}

\date{\today}

\maketitle

\section{Introduction}
Nonlinear light propagation in the atmosphere is of great interest as it can export the high intensity optical fields from the laboratory to the real world with applications
of remote sensing and atmospheric studies. Since  there is no optomechanical device that can control the properties of the beam
as it propagates freely through the atmosphere, the understanding of light matter interaction is an essential
tool to control radiation, intensity and profile of the beam in space.
Light filaments~\cite{Braun95,Zhao95} contain confined  regions of high intensity and can trigger various nonlinearities
leading to  supercontinuum and terahertz generation as well as strong field ionization and stimulated emission.
The latter mechanism is of particular interest because it can lead to forward and backward
air lasing~\cite{Luo03}.  \\
In this context there has been a particular interest in the
mechanism of stimulated emission of the molecular nitrogen ion itself.
Various mechanisms for gain are proposed; electron re-collision~\cite{Liu15},
strong field ionization~\cite{Yao16}, superradiance~\cite{Li14b} and strong field
interaction with plasma~\cite{Richter17}.

In this paper optical gain between
the vibrational ground states of  N$_2^+$ is analyzed with
high spectral and temporal resolution.  The time evolution of the plasma
following a 800 nm pump
is indirectly measured through stimulated emission driven by an ultrashort  delayed broadband seed around 400 nm.
The transitions
of interest (first negative band) are between the electronic states
of  $B^2\Sigma^+_g$ (upper state) and the $X^2\Sigma_g^+$ (lower state),
both in the ground state of vibrational energy ($\nu=0$). The ultrafast dynamics of
emission between individual ro-vibrational states suggests that two nonlinear processes
of resonant stimulated Raman scattering and stimulated gain are working in parallel
resulting in an intricate emission spectrum that depends on the phase and population
of the relevant states. The coherence waveform generated by phase
locking of ro-vibrational states in upper and lower state combined
with the population modulation in stimulated Raman scattering results in a transient gain.

The fluorescence emission of $N_2^+$ emanating from levels with a lifetime of 62 ns~\cite{Wuerker88}
has been observed for centuries in Aurora Borealis.
Given the vibrational quantum number $\nu$ and the rotational
quantum number $J$, the  energy levels can be tabulated through:
\begin{eqnarray}
E&=&\omega_e(\nu+\frac{1}{2}) -X_e \omega_e (\nu+\frac{1}{2})^2  +Y_e \omega_e(\nu+\frac{1}{2})^3  \nonumber \\
&+& B_{\nu}  J(J+1) -D_{e} J^2(J+1)^2 +T_e.
\label{eq:lines}
\end{eqnarray}

The first three terms are vibrational, the fourth and fifth terms are rotational,
and the last term is the energy level of the electronic state for $J=0$ and $\nu=0$.
The subscripts $\nu$ and $e$ refer to vibration and electronic states.

The values for selected levels
are (in cm$^{-1}$) $ \nu'=0$, $T_e = 0$, $\omega_e = 2207.00$, $\omega_e  X_e$ = 16.10, $\omega_e Y_e$ = -0.040, $\alpha_e=0.01881$, $B_e$ = 1.93176, $D_e = 6.1\cdot 10^{-6}$
for the {\bf X} state (ion ground state)
and $T_e$ = 25461.4, $\omega_e$ = 2419.84, $\nu''=0$, $\omega_ e X_e$ = 23.18, $\omega_e Y_e$ = -0.537, $\alpha_e=0.024$, $B_e$ = 2.07456, $D_e = 6.17\cdot 10^{-6}$
for the {\bf B} state~\cite{Klynning82}.
$B_{\nu}$ is calculated as $B_{\nu}= B_e - \alpha_e(\nu+\frac{1}{2})$.
The selection rules for transitions between rotational states are
$\Delta J = 0, \pm 1$.
Since both electronic states under consideration are $\Sigma$ states,
$\Delta J=0$ ({\bf Q} branch) is forbidden. The possible transitions from  ${\bf X}$ to ${\bf B}$ state are labeled with two branches of {\bf P} (lower energy) and {\bf R} (higher energy). Historically  the  spectrum was taken for absorption spectroscopy and the final state is
usually referred to as the state with higher energy, so $\Delta J=1$  results in higher energy photons and $\Delta J=-1$ in lower energy.

A non-resonant ultrashort pulse excites the rotational states in neutral and ionized molecules.
The pulse length is much shorter than the characteristic revival times of $1/(2Bc)$, where $B$
is the rotational constant. Therefore the pulse spectrum covers the full  bandwidth of rotational energies in neutral and cation states.
In non-adiabatic regime of alignment with ultrashort pulses~\cite{Stapelfeldt03}, the
 molecular response is too slow to follow the rapidly changing potential.  The impulse like
 excitation imparts
 a fixed phase relation on the rotational states.
 All diatomic molecules (including those that are ionized by
 the high field), are given a torque which, quantum mechanically,
 modifies the $J$ distribution through  Raman transitions with $\Delta J=\pm 2$.
    Since this Raman excitation proceeds via  discretely spaced levels at a well
localized initial time, a ``wavepacket'' is created~\cite{Varma09}.
The discrete frequencies separated by $4Bc$  will ``rephase''
at time intervals of $1/(2Bc)$,  a process similar to phase locking of the modes
in a laser cavity resulting in
 the generation of ultrashort pulse trains.  The restoration of wave packets
 at equal time intervals referred to as ``revivals''~\cite{Yeazell91} takes place as long
 as the phase coherence is maintained.
The change of alignment induced
 by linearly polarized light can be simply monitored by $\langle \cos \theta(t)^2 \rangle$,
 where $\theta$ is the angle between
 the molecular axis and the light polarization averaged over the ensemble of molecules.
 The time evolution of alignment of molecular ensemble is
 monitored through transient birefringence resulting on the wavelength shift
 of the probe pulse \cite{Marceau10}.

\section{Experimental setup}

A train of 50 fs pulses at 1 kHz, 1.3 mJ energy  each, centered at 795 nm,
from a Ti:sapphire oscillator regenerative amplifier (Coherent-Hidra),  is focused (N.A 0.01) into a long cell filled  with nitrogen.
A probe pulse is created by frequency doubling  10\%  split-off the main beam.   A dichroic mirror is used to overlap the 795 nm pump beam with its second harmonic.  The 940 $\mu$J pump and 140 nJ probe are focused into a nitrogen cell.
 The polarization of each beam is controlled with appropriate  wave plates.
 The transmitted beam is collected two meters from the geometrical focus,
 and  focused onto the entrance slit of a spectrometer. The spectra are acquired with
 a 1 meter Czerny Turner spectrometer  HR 640  from Instruments SA Inc with reciprocal line dispersion of 12 $A^0 /$mm.
The 795 nm beam  ionizes the medium, while the second harmonic beam is used to seed
the   {\bf X}~$(\nu=0) \leftarrow$ {\bf B}~$(\nu=0)$ transitions at 391 nm.
The delay between probe and pump is stepped with 7 fs increments.  At each step the spectrum of the N$_2^+$ emission is measured over a 4 nm bandwidth with a
resolution of 2 cm$^{-1}$.

\begin{figure}[h!]
    \centering
\includegraphics*[width=\linewidth]
{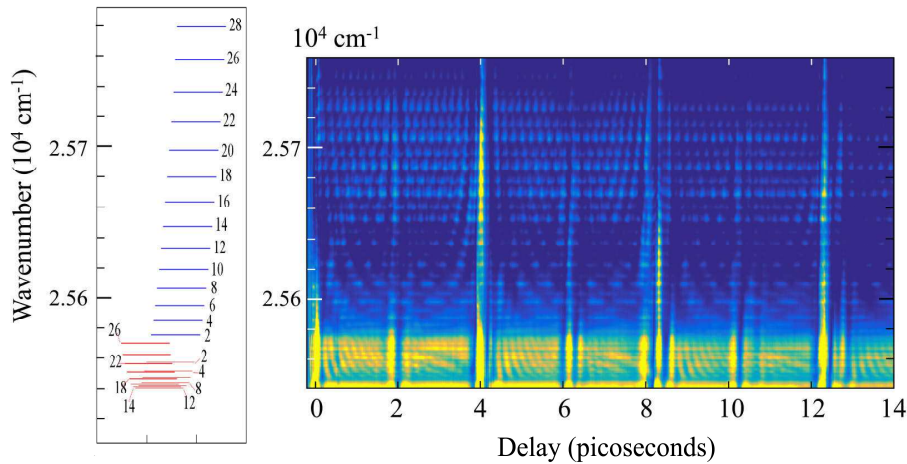} \caption{\small Left: energy diagram of the P branch
(in red) and the R branch (in blue). The
wavelengths are indicated on the left scale in cm$^{-1}$. The lines are calculated
from reference~\cite{Klynning82}.   Right: The spectra
of the  emission from {\bf X} ($\nu=0) \leftarrow$ {\bf B}($\nu$=0) as a
function of pump probe delay; time dependent gain is visible as a function of delay between pump
at 795 nm(940$\mu$J) and seed at 400 nm(140 nJ) for pure nitrogen at 100 torr. The color coding shows the intensity increasing from blue to red, the emission data being presented in logarithmic scale.
 Each individual rotational transition, inhibits two time dependent modulation. An almost inphase oscillation occurs for all transitions  at times corresponding to multiples of half revival. Each rottaional line undergoes oscillation with shorter period that increases with the rotational number. }
\label{fig:3Dspectrum}
    \end{figure}

A spectrogram is shown in Fig.~\ref{fig:3Dspectrum}.
It should be noted that no emission is observed in the absence of probe pulse,
which implies that the radiation is associated with a gain
mechanism rather than a spontaneous emission or fluorescence.
This confirms also that there is not enough spectral broadening of the pump
to create self seeding.

The spectrum of  N$_2^+$ emission  ${\bf X}  (\nu=0) \leftarrow {\bf B} (\nu=0)$  is shown in Fig.~\ref{fig:3Dspectrum} for each delay of the second harmonic seed following the 795 nm pulse. There is a clear delay
 in the order of 20 fs between the zero delay and observation of ro-vibrational transitions. The spectrogram
at each delay is a vertical strip of the image.   The emission is integrated for hundreds of miliseconds which is orders of magnitude longer than the lifetime of the emission. The time resolution on this measurement is solely due to resolution in time delay between the center of the two ultrashort pulses. The time dependent gain is due to the evolving plasma generated by 795 nm pulse. The brightness of emission as a function of delay creates parabolic lines that are
symmetric around the multiples of half revivals (Fig.~\ref{fig:3Dspectrum}).  Some selected transitions
are shown on the  Fig.~\ref{fig:3Dspectrum}.\\

Each horizontal line in  Fig.~\ref{fig:3Dspectrum} shows the time evolution of the emission as a function of pump-probe delay for individual $P_J$ or $R_J$.
It is clear that all the emissions undergo a strong intensity modulation
at the time corresponding to multiple of half revivals of $N_2^+$.
The strong modulation at revivals seen as dark and bright vertical
lines are recorded  for delays up to 150 ps (limited by the range of translation stage).
 The fast modulation at each line (visibly increasing with angular momentum $J$ in  the R branch)
 is due to stimulated resonant Raman emission that couples the state $J$ to adjacent
 rotational states that are spaced by multiples of $|\Delta J|=2$.

\section{Analysis and discussion}

It is a common practice to present the time dependent  gain by
integration over groups of transitions, mostly due to limitation in resolving power
of the spectrometer~\cite{Zhang13}.  Even with our spectral resolution
that identifies most of the emissions in R  and some in the P branch,
one can gain physical insight into the gain dynamics from such presentation.
\begin{figure}[h!]
    \centering
\includegraphics*[width=\linewidth]
{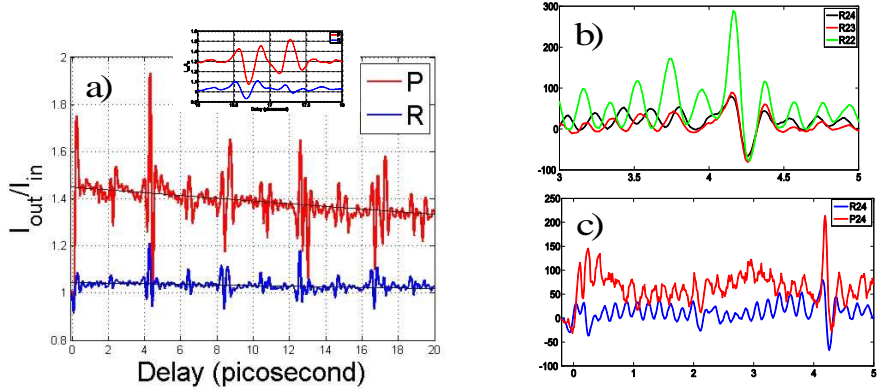} \caption[]{\small (a)   Gain in {\bf P} and  {\bf R} branches
 of emission corresponding to integration of all transitions
     from {\bf B} to {\bf X} states with $\Delta J=-1 $ and $\Delta J =+1$ as a function of delay between
     pump and seed. An inset shows the time
     evolution between 16 and 18 ps. The oscillation is visible in both {\bf P} and {\bf R} branch
     but they are not completely in phase. (b) Time evolution of the gain for
     {\bf R } transition from rotational states of 22, 23 and 24. (c) Time dependent emission for $R_{24}$ and $P_{24}$ that originate from the same upper state.  }
\label{fig:prtime}
    \end{figure}
 Figure~\ref{fig:prtime} shows the time dependent emission in integrated P and R branch (a)
 and from individually selected rotational lines (b) as a function of seed delay.
 An abrupt gain is visible as the seed follows the pump laser.
 The signal for both {\bf P} and {\bf R} emission decays exponentially
 as a function of seed delay; {\bf P} branch signal decays to $1/e$ of its maximum after 240 ps,
 whereas  the {\bf R} branch decays in 800 ps for nitrogen at 100 torr. Note that the decay time for these two channels are significantly different. 
The exponential decay is fitted to the data (black lines in Fig.~\ref{fig:3Dspectrum} a).
Both {\bf P} and {\bf R} emissions are modulated at revivals and fractional
revivals with the oscillation duration increasing at larger time delays with
lower modulation depths.

The oscillations of the integrated {\bf P} and {\bf R} emissions
are not quite in phase as shown in the inset of Fig.~\ref{fig:prtime}(a); the
 oscillations start nearly in phase at the first modulation and change  to
180 degree out of phase after a picosecond.
The emission from selected lines at the far end of spectrum from  ${\bf R}$ lines
are shown in [Fig.~\ref{fig:prtime}(b)].
A collective enhancement of gain followed by absorption is observed  at half revival.
This coherence behavior has been referred to as a superradiance effect ~\cite{Dicke54}
by recent papers \cite{Li14b,Lei17}. 


The gain at each ro-vibrational transition depends on the stimulated emission cross section. The probability of emission is proportional to
the dipole moment $<\Psi_{J,B}|\mu. e|\Psi_{J-1,X}>$, which, for an   ${\bf R}$ transition from upper state ``J''  has a time dependent
 phase of $2 \pi c[ (B_BJ(+1)-B_XJ(J-1))]t$.  The ionizing pulse sets the time zero for the rotational states, then each evolves based on
the rotational constant  and angular momentum of the level.
The emission stimulated by the probe  undergoes maximum and minimum based on the relative phase of the connecting states.\\

\begin{figure}
    \centering
\includegraphics*[width=\linewidth]
{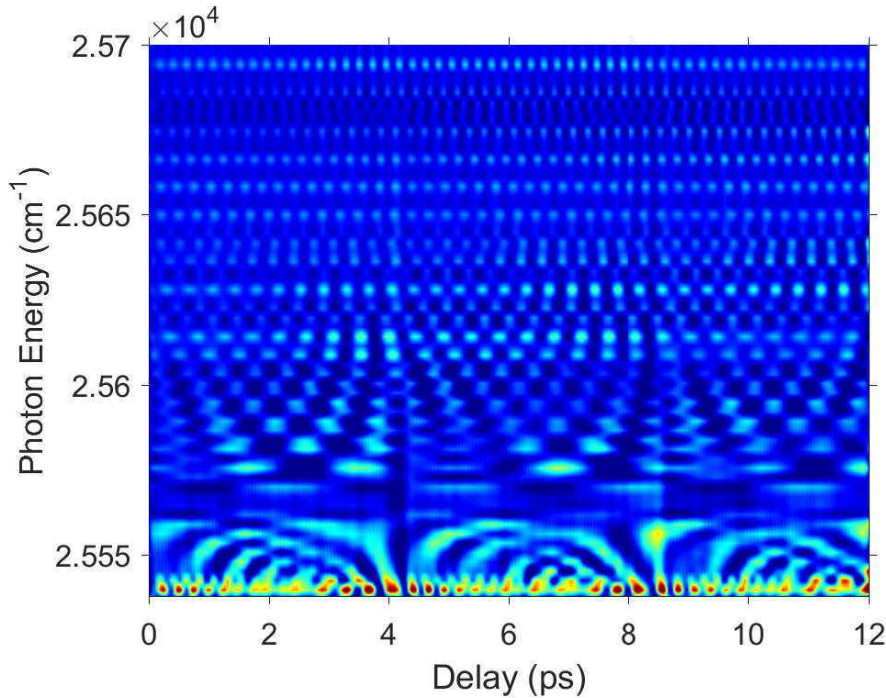}\caption{\small  Simulation of the time dependent gain, presented by calculation of imaginary part of susceptibility.}
\label{fig:simulation}
\end{figure}

In order to have a better understanding of time dynamics of the emission, we  made a simulation using time dependent Schrodinger equation. In our model  two electronic levels of N$_2^{+}$ were considered: {\bf X}($\nu$=0) and {\bf B}($\nu$=0) with a dipole transition independent of the rotational states.  Rotational states were uniformly populated at time zero from J=0 to J=20, with phases set to zero.  The relative populations were set to P({\bf X}) = 0.7 and P({\bf B}) = 0.3.  A probe pulse with an intensity of $4\times 10^{10}$ W/cm$^{2}$ and a duration of 24 fs caused electric dipole coupling between the states and redistributed population depending on the relative phases of the states.  The total dipole moment of the system versus time was calculated for 12 ps, from which the linear susceptibility of the medium in the frequency domain was calculated.  The gain or loss of the probe spectrum was calculated from the imaginary part of the susceptibility (Fig. \ref{fig:simulation}). The simulation resembles the experimental observation of time dependent gain at each individual transition. the collective effect of in phase emissions at revivals are not included in the model.

The nearest Raman transitions impart a population modulation upon each state ``J'' proportional to
 $ \alpha \exp (2 i \pi c B_\ell(J+2)(J+3)t) + \beta  \exp (2 i \pi c B_\ell(J-2)(J-1)t)$,  where $\alpha$ and $\beta$
 are probabilities of Raman transitions for $\Delta J =2$ and $\Delta J =-2$, respectively.
 The oscillation frequency on each rotational transitions is increasing with the  rotational
 state number ${\bf J}$ as shown in  Fig.~\ref{fig:prtime}(b).

 At delays corresponding to  fractional revivals both absorption and emission are  observed (e.g  for half revival, emission is enhanced at 4.2 ps and absorption is recorded for 4.32 ps).  In other times, the emission
 from rotational lines is modulated over a zero floor and no absorption is recorded.

Comparing the seed amplification with a common upper state ${\bf B}  $ reveals the importance of
dynamics in the ground state ${\bf X}$. In Fig.~\ref{fig:prtime}(c), the two transitions from upper
state $J=24$ to lower state $J=23$ and $J=25$ result in two distinct lines of ${\bf R_{24}}$ and
${\bf P_{24}}$, respectively. It can be seen that at delays outside revivals the two lines
are 180 degree out of phase, a phase opposition that we attribute to the resonant stimulated Raman.
When the population is transferred from $J=25$ to $J=23$ in the $\bf X $ state, the
transition probability of  the {\bf R}$_{24}$ line
is enhanced at the expense of lowering the transition probability of
the {\bf P}$_{24}$. The nearly perfect out of phase relation in the emissions observed from the same upper state suggests that the gain is merely driven by the phase relation between the states.  Both lines experience  synchronised emission
and absorption at times corresponding to  alignment and anti-alignment.
The in phase emission is not perfect for all rotational states  as is
visually observed in Fig.~\ref{fig:3Dspectrum}.
 The modulation at  $\frac{3}{2}$ revival  (Fig.~\ref{fig:prtime} a inset) exhibits a synchronized brightness at first
enhancement which is followed by a nearly out of phase emission  in the two branches.
This suggests the effect of stimulated Raman in shifting the wavepacket population
in favor of $\Delta J=1 $ or $\Delta J =-1$.

Note that no exponential decay is observed from the emissions of individual lines,
a parameter that suggests a slowly decaying population inversion~\cite{Zhang13}.
Our measurements suggest that the relative phase
 between the rotational states has a major role in the amplification of the seed,
 dominating any possible contribution from population inversion.
\\

\begin{figure}[h!]
    \centering
\includegraphics*[width=\linewidth]
{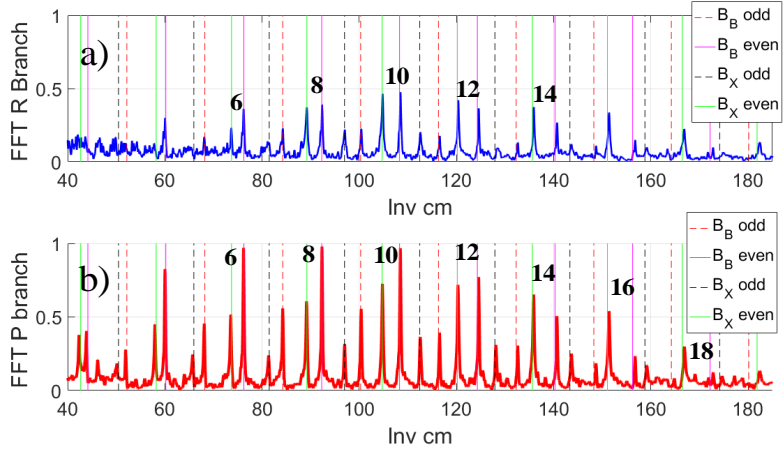}\caption{\small The Fourier amplitude of ${\bf R}$ (top) and ${\bf P}$ (bottom) branch emissions. The Fourier transform is plotted on a grid defined by Raman transitions of $\Delta J=-2$ in  ${\bf B}$ and  ${\bf X}$ state.
The distribution of  ${\bf X}$ is shifted to higher rotational states for both cases. The  ${\bf R}$  branch connects to lower states of  ${\bf X}$ compared to  ${\bf P}$. This result is in agreement with exchange of orbital angular momentum to higher ( ${\bf P}$) and lower  (${\bf R}$) values.}
\label{fig:Fourier}
\end{figure}
A qualitative representation of the rotational wavepacket distribution is obtained
through a  Fourier transform
of the emission lines in both  ${\bf P}$ and ${\bf R}$ sections of the spectrum.
The absolute value of the Fourier transform of integrated emission at two branches are presented in Fig.~\ref{fig:Fourier}  from a  nitrogen cell at 100 torr.
The Fourier transform of the ${\bf P}$ (red) and $ {\bf R}$ (blue) branches are presented separately in two plots with vertical grid lines corresponding to
the Raman transitions of $\Delta J=-2$ in the ${\bf B }$ branch and ${\bf X }$ branch.
Dashed lines correspond to upper states with odd rotational number
and solid lines correspond to even states. \\
The Fourier transform of the emission is well aligned with the grid of the Raman transitions.
The spin statistics of nitrogen leads to having the population ratio of two to one between even and odd states
of the thermal sample, which does not hold for the stimulated emission \cite{Azarm17}.
The Fourier transform reveals the signature of the wavepacket in both   ${\bf B }$  and  ${\bf X  }$ states, with a distinct shift in distribution.
Higher rotational states are present in the ground state as compared to the excited state.
This can be visualized by comparing the Fourier transform component of adjacent rotational lines in  ${\bf B }$  and  ${\bf X  }$.
For lower rotational numbers the amplitude of  ${\bf B }$  is higher than ${\bf X}$.
This trend is reversed  for rotational states higher than 10 in the ${\bf R }$ branch and 12 in the ${\bf P }$ branch. For example,
the  ratio of {\bf B} to {\bf X}  contribution  at rotational number 6 in the ${\bf P }$ branch  is 3/2,
while at rotational number 14 the same ratio  is 2/3 in the ${\bf R }$ branch emission.

The shift in the wavepacket distribution in ${\bf B }$ and ${\bf X }$ state
suggests that the two states have different interaction with ultrashort pulses.
Both distributions are shifted to higher rotational numbers
than a thermal distribution \cite{Arissian16}, with
${\bf X}$ state occupying the highest rotational numbers. By comparing the Fourier amplitude of ${\bf B }$ and ${\bf X}$ in lower rotational states one might conclude that the ${\bf B }$ state has a higher population that the ${\bf X }$ state,
hence the stimulated emission would be governed by a traditional population inversion.
Since the gain is observed for all the rotational states contributing in the observed
Raman process of Fig.~\ref{fig:Fourier}, independently
of the ratio of ${\bf B}/{\bf X}$ we conclude that the gain is
not determined by the population inversion but  coherence in the system. This coherence has been referred to as lasing without
inversion ~cite{Kocharovskaya86, Scully89}.
This gain could be further tailored by engineering the wavepacket
in the system with a two pulse alignment~\cite{Lee04} or a chirped seed counteracting the chirp in the level spacing of the rotational levels.\\

There are several observations in our  measurement
that reveal the major role of phase (time) dependent of the gain.
\begin{enumerate}
\item{ A delay of 20 fs
between the peak of the pulse and air lasing suggests that the lasing was delayed
until the wavepacket is formed in the ion states. (Similar delay is observed by
Li~\cite{Li14}).  A proper phase needs to be established between the rotational
states involved at each transition.}
\item{ The emission from individual line is measured on a
zero background with no exponential decay. This suggests that the lasing on each line
(in our experiment)
is strongly dependent on the phase of the lines rather than a population difference.}
\item{
The emission spectrum in air lasing shifts to lower rotational states at larger
delays as seen in Fig.~\ref{fig:3Dspectrum}. The coherence of rotational states
depends on the collision rate in the sample, with the higher rotational states
having longer coherence time. A decay of 64.5 $\pm5 ~ps/atm$ is measured for pure nitrogen at temperature of 295 K   ~\cite{Owschimikow10}, with $J=14$ having  50 percent longer delay as compared to  $J=4$.  Our observation shows that higher rotational states leave the gain process faster and
they no longer contribute
as an amplifier at revivals. For a sample at 100 torr the {\bf R} branch center shifts 2 cm$^{-1}$
to lower wavelengths for 20 ps delay. }
\item{
 The emission at individual lines as seen in Fig.~\ref{fig:prtime} are nearly in phase at revivals.
Gain and absorption are recorded for each line at the time corresponding to a half revival with a relative delay of 120 fs.
A  simillar delay is observed between alignment and anti-alignment of neutral nitrogen molecules under comparable laser intensities~\cite{Dooley03}.
It is known that ionization is enhanced for nitrogen molecules aligned with the laser field~\cite{Litvinuk03} hence the dipole transition between the ionic states depends on the alignment of the involved states.
The chirp in spacing of rotational energies results in phase delay between individual
lines and the sum over all states creates multiple oscillation as shown in inset of Fig.~\ref{fig:prtime}(a).}

\item{ The phase relation between
the rotational states will result in coherent population trapping, in which the wavepacket no longer
interacts with light. In this case any population in the upper state can make transitions to
the state orthogonal to the trapped one~\cite{kocharovskaya92}.
Absorption or gain will be observed depending on the phase of the coherence.}  \\
\end{enumerate}
A transient inversion is created with assistance of resonant stimulated Raman
scattering at delays outside revivals and the in phase dipole emission from rotational states in a wavepacket at
time of revival results in a collective behavior of the states with
applied field.  This collective behaviour could be due to  quantum interference of multiple excitation paths (previous item) or collective emission of emitters
known as superradience~\cite{Dicke54}.

\section{Conclusion}

We have performed  high resolution spectroscopy of  N$_2^+$ emission
${\bf X}  (\nu=0) \leftarrow {\bf B} (\nu=0)$  in a pump-probe setup,
using ultrashort pulses. By careful measurements of time dependent emission
from well identified rotational lines,
we can identify the  coherent mechanisms involved in
obtaining gain in nitrogen molecular ion in ultrashort pulse
high energy lasers. Resonant stimulated Raman scattering
 is responsible for the transient gain in the medium at all times.
 This effect is largely enhanced at the times corresponding to the
 rotational revival of the molecule, in which various rotational states are in phase. Our study suggests that coherent control of the rotational states in the molecular ion may result in control of the brightness of such emissions.

\section{Acknowledgement}
We had benefited from long discussions with Paul Corkum,
 Michael Spanner, Mathew Britton and Patrick
Laferriere from NRC and University of Ottawa.
This work is supported by the
U.S. Department of Energy (DOE) (DE-SC0011446), and the Army Research Office (ARO) (W911-NF-1110297).

\bibliography{c:/document/bib/ad,c:/document/bib/en,c:/document/bib/oz,c:/document/bib/refbook}
\end{document}